\documentclass[aps,prb,twocolumn,showpacs,superscriptaddress,floatfix,nofootinbib,10pt]{revtex4-1}

\usepackage{graphicx}
\usepackage{amsmath}
\usepackage{epsfig}
\usepackage{helvet}
\usepackage{amssymb}

\newcommand{\be}{\begin{equation}}
\newcommand{\ee}{\end{equation}}
\newcommand{\bea}{\begin{eqnarray}}
\newcommand{\eea}{\end{eqnarray}}

\newcommand{\tr}{\mbox{tr}}
\newcommand{\tra}{\mbox{\scriptsize tr}}
\newcommand{\zz}{\bar{z}}
\newcommand{\shr}{\mbox{\scriptsize shr}}
\newcommand{\bra}[1]{\mbox{$\langle #1 |$}}
\newcommand{\ket}[1]{\mbox{$| #1 \rangle$}}

\newcommand{\proj}[1]{\mbox{$|#1\rangle \!\langle #1 |$}}

\def\tr{ \mbox{tr}}

\begin{document}

\title{
Scaling of entanglement entropy \\in the (branching) multi-scale entanglement renormalization ansatz}
\author{G. Evenbly}
\affiliation{Institute for Quantum Information and Matter, California Institute of Technology, MC 305-16, Pasadena CA 91125, USA}
\email{evenbly@caltech.edu}
\author{G. Vidal}
\affiliation{Perimeter Institute for Theoretical Physics, Waterloo, Ontario N2L 2Y5, Canada}  \email{gvidal@perimeterinstitute.ca}
\date{\today}

\begin{abstract}
We investigate the scaling of entanglement entropy in both the multi-scale entanglement renormalization ansatz (MERA) and in its generalization, the branching MERA. We provide analytical upper bounds for this scaling, which take the general form of a boundary law with various types of multiplicative corrections, including power-law corrections all the way to a bulk law. For several cases of interest, we also provide numerical results that indicate that these upper bounds are saturated to leading order. In particular we establish that, by a suitable choice of holographic tree, the branching MERA can reproduce the logarithmic multiplicative correction of the boundary law observed in Fermi liquids and spin-Bose metals in $D\geq 2$ dimensions.
\end{abstract}


\maketitle

\section{Introduction} \label{sect:intro}

Understanding the collective behavior of quantum many-body systems has long presented a formidable challenge due to the exponential growth of Hilbert space dimension with system size $N$. In recent years, tensor networks \cite{Cirac09, Evenbly11a} have emerged as a formalism to efficiently describe some many-body wave-functions. By construction, tensor networks can only cover a small region of the huge Hilbert space, but this small region already seems to contain many states of interest, including the ground states of a large variety of many-body Hamiltonians.

\subsection{Tensor networks and the boundary law}

The boundary law\cite{Srednicki93,Latorre04,Plenio05,Bravyi06,Eisert06,Hastings07c,Masanes09,Eisert10} (also known as area law) for entanglement entropy has played --and continues to play-- a particularly important role in the development of tensor networks. Let us consider a many-body system defined on a lattice $\mathcal{L}$ and in a pure state $\ket{\Psi}$. The entanglement entropy of a region $\mathcal{B}$ of the lattice is defined as the von Neumann entropy $S(\rho)$ of the reduced density matrix $\rho$ for that region, where $S(\rho)$ and $\rho$ are given by
\begin{equation}\label{eq:s1entropy}
    S(\rho) \equiv -\tr [\rho \log_2 (\rho)],~~~\rho \equiv \tr_{\mathcal{L-\mathcal{B}}} \proj{\Psi}.
\end{equation}
 In a lattice $\mathcal{L}$ in $D$ spatial dimensions, we say that the state $\ket{\Psi}$ obeys a boundary law if the entanglement entropy of a region of $l^D$ lattice sites grows as $l^{D-1}$,
\begin{equation}\label{eq:s1scale}
    S_l \approx l^{D-1},
\end{equation}
i.e. if the entanglement entropy is proportional to the size $|\partial \mathcal{B}| \approx l^{D-1}$ of the boundary $\partial \mathcal{B}$ of the region, and not to the size $|\mathcal{B}| \approx l^{D}$ of the region itself. A highly non-trivial observation is then that the ground state of a generic local Hamiltonian happens to obey a boundary law for entanglement entropy, possibly with a multiplicative logarithmic correction\cite{Holzhey94,Callan94,Fiola94,Vidal03b,Wolf06,Gioev06,Li06,Barthel06,
Swingle10,Swingle09b,Motrunich07,Senthil08,Liu09},
\begin{equation}\label{eq:s1scale2}
    S_l \approx l^{D-1} \log_2 (l).
\end{equation}
For instance, Table \ref{tab:FreeFerm} summarizes the scaling of entanglement entropy in systems of free fermions\cite{Wolf06,Gioev06,Li06,Barthel06}. It shows that a free fermion ground state typically obeys the boundary law of Eq. \ref{eq:s1scale} except in the presence of a (sufficiently large) Fermi surface, in which case it displays a logarithmic correction as in Eq. \ref{eq:s1scale2}. More generally, there is abundant evidence, both theoretical and numerical, that the ground state of most many-body systems obey either a strict boundary law or a boundary law with a logarithmic correction.

The boundary law for entanglement entropy has had a major impact on the development of tensor networks in at least two ways. On the one hand, upon observing that the entanglement entropy of a \textit{generic} state of the many-body system scales instead according to a bulk law,
\begin{equation}\label{eq:s1scale3}
    S_l \approx l^{D},
\end{equation}
it became evident that ground states are very special, \textit{non-generic} states in the huge many-body Hilbert space. This realization implied that, while there is no efficient representation for generic states, one may still be able to find an efficient representation for ground states. Obtaining such an efficient representation has since then been the goal of the tensor network program.

\begin{table}[!bth]
\begin{tabular}{|c||c|c|c|c|}
\hline
                 & Gapped & \multicolumn{2}{|c|}{Gapless}\\
                 &                &   $D-D_\Gamma \geq 2$   & $D-D_\Gamma=1$        \\  \hline
  $\ D=1\ $      &   const.       &    -                  & $\log_2 (l)$            \\
  $\ D=2\ $      &   $l$          &    $l $               & $\ l \log_2 (l)\ $      \\
  $\ D=3\ $      &   $l^2$        &    $l^2$              & $\ l^2 \log_2 (l)\ $    \\
\hline
\end{tabular}
\caption{Scaling of the entanglement entropy $S_l$ of a block of $l^D$ sites in the ground state of a free fermion system. The scaling of entanglement entropy depends only on the spatial dimension $D$, and on general properties of the low energy spectrum of the system. In particular, on whether there is an energy gap and, in the gapless case, on the difference $D-D_\Gamma$ between the spatial dimension $D$ and the dimension $D_{\Gamma}$ of the Fermi surface. All cases are examples of Eqs. \ref{eq:s1scale} or \ref{eq:s1scale2}.}
\label{tab:FreeFerm}
\end{table}

On the other hand, over the last ten years the boundary law has served as a main guiding principle when designing new tensor networks. The rationale is simple: in order to build a tensor network for a particular class of ground states, one must first ensure that the tensor network is capable of reproducing the corresponding scaling of entanglement entropy. Let us mention three prominent examples.

The first example refers to the proposal of the \textit{multi-scale entanglement renormalization ansatz}\cite{Vidal08} (MERA) to represent ground states of quantum critical (and thus gapless) systems in $D=1$ spatial dimensions. The \textit{matrix product state}\cite{Fannes92,White92} (MPS), historically the first tensor network and responsible for the impressive success of the density matrix renormalization group\cite{White92}, is able to reproduce the boundary law (i.e., constant entanglement entropy $S_l$) characteristic of gapped systems in one dimension. However, the MPS cannot reproduce the logarithmic scaling $S_l \approx \log_2(l)$, characteristic of gapless/quantum critical systems in one dimension. That motivated the proposal of an alternative tensor network, the MERA, for the purpose of obtaining an efficient tensor network representation that, in $D=1$ dimensions, could reproduce a logarithmic scaling of entanglement entropy\cite{Vidal08b} and thus describe gapless/quantum critical systems.

A second example refers to the proposal of \textit{projected entangled-pair state}\cite{Verstraete04} (PEPS) and MERA in two and larger spatial dimensions. In $D>1$ dimensions, the MPS can not reproduce the ($D$-dimensional) boundary law of Eq. \ref{eq:s1scale}. Accordingly, an MPS can not be used to efficiently describe ground states on large systems in $D>1$ dimensions (and yet, combined with finite size scaling techniques, they have proven useful as a numeric tool for investigating some such systems; see for instance Ref. \onlinecite{2DMPS}). PEPS and MERA where then proposed as tensor networks that naturally reproduce the boundary law in $D>1$ dimensions. As such, they are plausible candidates to describe the ground state of gapped systems and of some gapless systems in $D>1$ dimensions, see the first and second columns of Table \ref{tab:FreeFerm}.

The third example is given by the recent proposal of the branching MERA\cite{Evenbly12a, Evenbly12b}, whose properties are further investigated in this paper. As mentioned above, PEPS and MERA can reproduce the boundary law in $D>1$ dimensions. However, they cannot surpass it. This means that ground states with a logarithmic violation of the boundary law in $D>1$ dimensions, such as those of Fermi liquids and spin-Bose metals\cite{Motrunich07}, cannot be efficiently described by these tensor networks. The branching MERA has been proposed to overcome this limitation of PEPS and MERA. To each realization of the branching MERA one can attach a holographic tree that describes its branching structure in scale space\cite{Evenbly12b}. In particular, for certain choices of branching in the holographic tree, one can then reproduce the logarithmic correction to the boundary law, Eq. \ref{eq:s1scale2}, in $D>1$ spatial dimensions. Thus the branching MERA is a plausible candidate to efficiently describing the ground state of Fermi liquids and spin-Bose metals.

In summary, as illustrated in Table \ref{tab:TensorNetworks}, we now have efficient tensor networks capable of reproducing all the known types of scaling of entanglement entropy in the ground state of a local Hamiltonian, Table \ref{tab:FreeFerm}.

\begin{table}[!htb]
\begin{tabular}{|c||c|c|c|c|}
\hline
                 &       Gapped         & \multicolumn{2}{|c|}{Gapless}\\
                 &   $S_l\approx l^{D-1}$ &   $S_l \approx l^{D-1}$  & $S_l \approx l^{D-1}\log(l)$  \\  \hline
  $\ D=1\ $      &   MPS                &    -                  & MERA            \\
  $\ D=2\ $      &   PEPS          &    MERA          & branch. MERA    \\
  $\ D=3\ $      &   PEPS          &    MERA          & branch. MERA    \\
\hline
\end{tabular}

\caption{Tensor networks for each known type of entanglement scaling. This table refers to the \textit{natural} scaling of entanglement in each tensor network, as obtained for large $l$ in an infinite system, assuming a random choice of variational parameters and fixed bond dimension for its tensors. [We emphasize that in practice, one can use the MPS for all types of systems, although not in a scalable way. Moreover, through proper fine-tuning of variational parameters, one can use PEPS to represent certain gapless systems in $D>1$ dimensions, and MERA to represent gapped systems in all dimensions.]}
\label{tab:TensorNetworks}
\end{table}

\subsection{Summary of results and structure of the paper}

The goal of this paper is to provide a detailed study of the scaling of entanglement entropy both in the MERA and in the branching MERA.

The scaling of entanglement entropy in the MERA was already derived in Ref. \onlinecite{Vidal08b}, where this tensor network was proposed. Unfortunately, the result was left outside the final published version, Ref. \onlinecite{Vidal08}, due to lack of space. Here we will start by reviewing and expanding on the original derivation of Ref. \onlinecite{Vidal08b} (see also Ref. \onlinecite{Barthel10}), which argued that the entanglement entropy in the MERA was upper bounded as $S_l \leq \log_2(l)$ in $D=1$ dimensions; and as $S \approx l^{D-1}$ in $D>1$ dimensions.

Re-deriving the scaling of entanglement entropy in the MERA serves the purpose of introducing several of the concepts and techniques that are needed for the main result of the paper: providing strict upper bounds on the scaling of entanglement entropy in the branching MERA. Indeed, these upper bounds are obtained by suitably generalizing, to the case of non-trivial holographic branching, the counting arguments used in the MERA.

In addition, for several cases of interest, we use numerics to show that the upper bounds for the scaling of entanglement entropy in the branching MERA are saturated by a random choice of variational parameters. In practice, we had to resort to the formalism of free fermions so that the required computations, involving large amounts of entropy, can be made affordable. However, as it is the case for the MERA, we expect that the saturation of the upper bounds in the branching MERA occurs generically, and not just for states of free fermions.

\begin{table}
\begin{tabular}{|c||c|c|c|c|}
\hline
                 &  \multicolumn{3}{|c|}{Branching MERA}\\
                 &   $b<2^{D-1}$        &   $b=2^{D-1}$ ${}^{[iii]} $         & $b=2^D$  ${}^{[iv]}$ \\ \hline
  $\ D=1\ $      &    ~~~     const.${}^{[i]}$            &   ~~~~$\log_2 (l)$ ${}^{[ii]}$       &   $l$     \\
  $\ D=2\ $      &   ~~~~$l$ ${}^{[ii]}$     &   $\ l \log_2 (l)$    & $l^2$     \\
  $\ D=3\ $      &   ~~~~$l^2$ ${}^{[ii]}$    &   $\ l^2 \log_2 (l)$  & $l^3$     \\

\hline
\end{tabular}
\caption{Scaling of entanglement entropy in the branching MERA, as a function of the space dimension $D$ and branching ratio $b$. Only a subset of possible branching MERAs is represented in this table. $[i]$ A constant entropy in $D=1$ is obtained with a branching ratio $b=0$, corresponding to a finite correlated MERA. $[ii]$ In all dimensions $D$, the choice $b=1$ corresponds to the regular (i.e. non-branching) MERA. $[iii]$ The choice $b=2^{D-1}$ produces a logarithmic correction to the $D$-dimensional boundary law. $[iv]$ For $b=2^D$ one obtains the bulk law characteristic of generic states in the many-body Hilbert space.}
\label{table:BranchMERA}
\end{table}

Let us summarize a subset of our findings, by focusing on a particularly simple subclass of branching MERAs, characterized by a branching ratio $b$ (to be further explained in the main text).
For $b=1$, which corresponds to the regular (i.e. non-branching) MERA, one recovers of course the scaling of the MERA described above. The same scaling is seen to also hold for any branching ratio $b < 2^{D-1}$.
For $b=2^{D-1}$, the branching MERA reproduces the logarithmic correction of Eq. \ref{eq:s1scale2}. This is, arguably, the most interesting case, which makes the branching MERA a potential candidate to describe strongly entangled ground states such as those of Fermi liquids and spin-Bose metals in $D>1$ dimensions.
Finally, and somewhat intriguingly, for $b=2^D$ (which is the largest possible value of the branching ratio $b$) we find that the branching MERA obeys the bulk law of Eq. \ref{eq:s1scale3}, thus matching the scaling of entanglement entropy observed in generic states of the many-body Hilbert space.

Table \ref{table:BranchMERA} summarizes these results. In addition, for values of $b$ in the range $2^{D-1} < b < 2^{D}$, as well as for other subclasses of branching MERA, on finds plenty of other forms of scaling, interpolating between the boundary and bulk laws.

The rest of the paper is divided into six more sections. Sects. \ref{sect:fund} and \ref{sect:entscale} review the relevant material on the MERA and derive upper bounds for the scaling of entanglement entropy in the MERA, respectively. Similarly, Sects. \ref{sect:fundbranch} and \ref{sect:branchbound} review the relevant background material for the branching MERA and derive upper bounds for the scaling of entanglement entropy in the branching MERA. Sect. \ref{sect:numresults} presents a numerical study, using free fermions, that shows the saturation of the upper bounds derived for several instances of the branching MERA. Finally, Sect. \ref{sect:conclusion} contains some conclusions.

\begin{figure}[!tbh]
\begin{center}
\includegraphics[width=8cm]{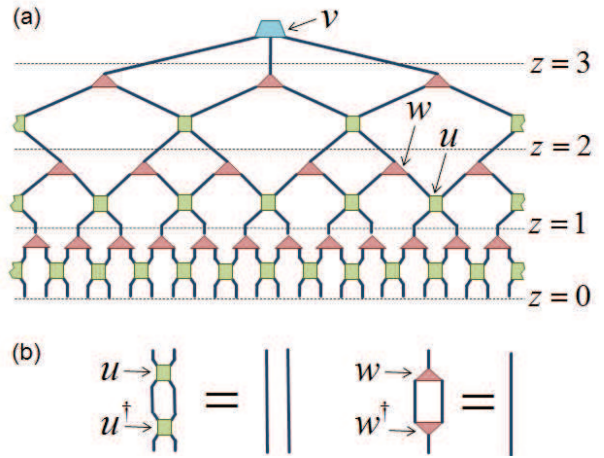}
\caption{(a) MERA for a lattice of 24 sites in $D=1$ dimensions. It consists of disentanglers $u$ and isometries $w$, as well as of a top tensor $v$ which will not be important in this paper. (b) Graphic representation of the isometric constraints imposed on disentanglers $u$ and isometries $w$, see also Eq.\ref{eq:s2e2}.} \label{fig:Intro}
\end{center}
\end{figure}

\section{Fundamentals of the MERA} \label{sect:fund}

Let us start by reviewing the basics of the MERA required in order to derive an upper bound on the scaling of entanglement entropy. Here we briefly describe the tensor network, the structure of its causal cones, and the computation of reduced density matrices.

\subsection{Tensor Network}

The MERA\cite{Vidal08} is a general class of tensor networks for quantum systems on a $D$-dimensional lattice. It is based upon a coarse-graining transformation known as entanglement renormalization\cite{Vidal07} (ER). For concreteness, in this manuscript we consider a hypercubic lattice $\mathcal{L}$ in $D$ dimensions (square lattice in $D=2$ dimensions, cubic lattice in $D=3$ dimensions, etc) and a particular MERA scheme based on the coarse-graining of a hypercubic block of $2^D$ lattice sites into a single lattice site, as illustrated in Fig. \ref{fig:Intro} for $D=1$ dimensions. \cite{Other}


More specifically, if $L^D$ is the total number of sites of the hypercubic lattice $\mathcal{L}$, then the MERA consists of $Z \approx \log_2 (L)$ layers of tensors, where each layer is in turn made of a sublayer of tensors called disentanglers $u$ and a sublayer of tensors called isometries $w$, see Fig. \ref{fig:Intro}(a). We parametrize the depth within the MERA in terms of the scale parameter $z$, where $z=0$ corresponds to the $L^{D}$ open indices of the MERA, i.e. the indices related to the $L^{D}$ sites of lattice $\mathcal{L}$; $z=1$ labels the $(L/2)^{D}$ indices connecting the first and second layer of tensors in the MERA; etc. We refer to $z$ as the scale parameter because an index connecting the layers $z-1$ and $z$ can be interpreted to describe an effective site that contains approximately $2^z$ sites of $\mathcal{L}$.


The disentanglers $u$ and isometries $w$ enact local mappings of the Hilbert space. A disentangler is a unitary transformation on a hypercubic block of $2^D$ sites, while an isometry defines a coarse-graining map from a block of $2^D$ sites into a single (coarse-grained) site. We assume a uniform bond dimension $\chi$ through the tensor network, which implies that each index can be viewed as representing a vector space $\mathbb{V}$ of dimension $\chi$. Then, disentanglers and isometries are maps
\begin{equation}
u:\mathbb V^{ \otimes 2^D}  \mapsto \mathbb V^{ \otimes 2^D} ,\; \; \; w:\mathbb V  \mapsto \mathbb V^{ \otimes 2^D}, \label{eq:s2e1}
\end{equation}
subject to the isometric constraints
\begin{equation}
u^\dag  u = \mathbb I^{\otimes 2^D} ,\; \; \; w^\dag  w = \mathbb I \label{eq:s2e2}
\end{equation}
where $\mathbb I$ is the identity operator on $\mathbb{V}$, see also Fig\ref{fig:Intro}(b). [In addition, a disentangler automatically satisfies $u u^\dag = \mathbb I^{\otimes 2^D}$.]

\begin{figure}[!tbh]
\begin{center}
\includegraphics[width=8cm]{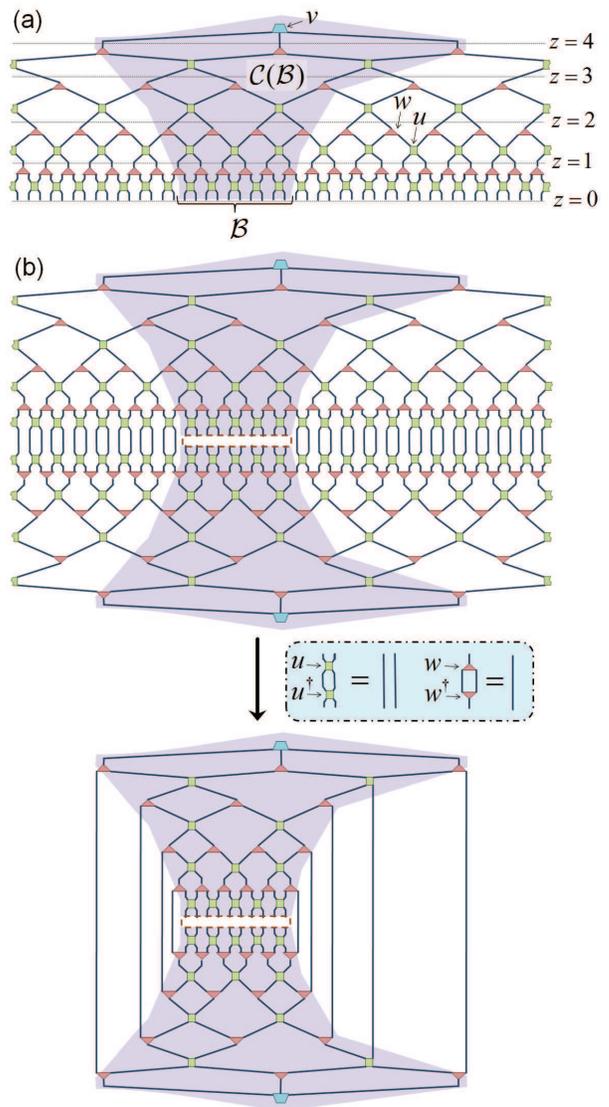}
\caption{(a) MERA representing a state $\ket{\Psi}$ on a lattice of $48$ sites in $D=1$ dimensions. Here the causal cone $\mathcal C(\mathcal B)$ of a block $\mathcal B$ of sites is shaded. (b) The reduced density matrix $\rho_{\mathcal B}$ can be obtained from $\ket{\Psi} \bra{\Psi}$ by tracing out all sites in the complement of $\mathcal B$, Eq.\ref{eq:s1entropy}. Due to the unitary/isometric tensor constraints, Eq. \ref{eq:s2e2}, many tensors annihilate to identity with their respective Hermitian conjugates, and only tensors within the causal cone $\mathcal C(\mathcal B)$ of block $\mathcal B$ remain.} \label{fig:1DCausal}
\end{center}
\end{figure}

\subsection{Causal cones} \label{sect:causal}


The tensor network corresponding to the reduced density matrix $\rho$ for a region $\mathcal B$ of sites is such that many of the tensors in the MERA annihilate to identity with their complex conjugates, see Fig. \ref{fig:1DCausal}. As a result, $\rho$ is actually a function of only a subset of the tensors in the MERA. We refer to this subset as the causal cone $\mathcal C(\mathcal B)$ of region $\mathcal B$ \cite{Vidal08}. Causal cones in the MERA have a characteristic form, resulting from the peculiar (discrete) geometry of the tensor network, that we now examine, starting with a lattice $\mathcal{L}$ in $D=1$ dimensions.


Let $\mathcal B$ be a region of $l_0$ contiguous sites of lattice $\mathcal L$, where we assume that $l_0$ is smaller than half the lattice size $L$. We denote by $l_z$ the \emph{width} of the causal cone at depth $z$, which is defined as the number of effective sites contained within the causal cone $\mathcal C(\mathcal B)$ at depth $z$. For instance, in Fig. \ref{fig:1DCausal}(a), the width of the causal cone $l_z$ as a function of the scale $z$ is $l_0=10$, $l_1 = 6$, $l_2=4$, $l_3 = 3$, and $l_4=3$.

\begin{figure}[!tbh]
\begin{center}
\includegraphics[width=8cm]{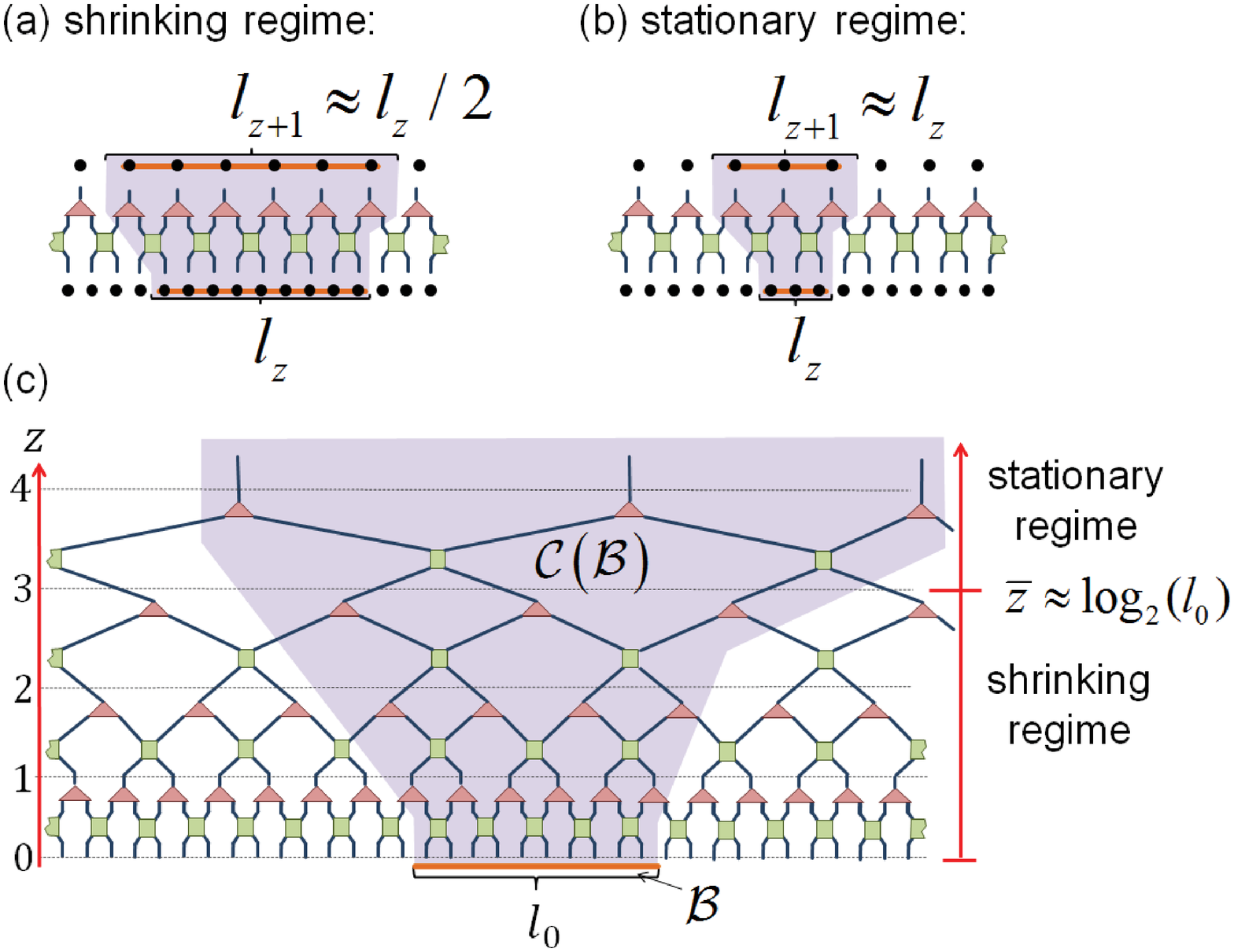}
\caption{(a) A region made of $l_z \gg 1$ sites is coarse-grained to a smaller region made of $l_{z+1} \approx l_z /2$ sites (shrinking regime). (b) A region with $l_z = 3 $ sites is coarse-grained into a region of equivalent width, i.e. $l_{z+1} = l_z = 3$ (stationary regime). (c) The width of the causal cone $\mathcal C(\mathcal B)$ of a block $\mathcal B$ comprised of $l_0 \gg 1$ sites shrinks with increasing scale $z$ until the crossover scale $\zz \approx \log_2(l_0)$ is reached, after which it remains stationary.}  \label{fig:CausalStructure}
\end{center}
\end{figure}

In order to understand the evolution of the width $l_z$ with the scale $z$, consider the action of a single layer of the MERA. The disentanglers $u$ act to spread the width of the causal cone by at most two sites (at each of the two boundaries of the region, left and right, there is a disentangler that can at most spread the causal cone by one site). On the other hand, the isometries $w$ act to compress the width of the causal cone by roughly a factor of two, see Fig. \ref{fig:CausalStructure}. As a result, the width $l_{z+1}$ of the causal cone at depth $(z+1)$ can be seen to be bounded by $l_z$ as follows,
\begin{equation}
\frac{l_z+2}{2} \le l_{z+1} \le \frac{l_z+4}{2}. \label{eq:s2e5}
\end{equation}
If the width of the causal cone at depth $z$ is much greater than one, $l_z \gg 1$, then under a layer of coarse-graining the action of the isometries dominates and the width shrinks by roughly a factor of two, $l_{z+1} \approx l_z /2$, see Fig. \ref{fig:CausalStructure}(a). We refer to this as the \emph{shrinking regime} of the causal cone. Conversely, if the width is much closer to one, $l_z = 2,3$ or $4$, then the spread of the support from the disentanglers could be exactly balanced by the shrinking of the support from the isometries, such that the causal cone could remain at a fixed width, i.e. $l_{z+1} = l_z $. We refer to this as the \emph{stationary regime} of the causal cone, see also Fig. \ref{fig:CausalStructure}(b).

The previous discussion implies that, as a function of the scale $z$, the causal cone $\mathcal C(\mathcal B)$ of a region $\mathcal B \in \mathcal L$ made of $l_0 \gg 1$ sites at $z=0$ experiences two regimes. From the smallest scale $z=0$ all the way up to some crossover scale $\zz \approx \log_2 (l_0)$, the causal cone is in the shrinking regime, whereas for scales $z$ larger than $\zz$, it is in the stationary regime, see Fig. \ref{fig:CausalStructure}(c).

On a $D-$dimensional lattice $\mathcal{L}$, the causal cones of the MERA behave similarly. If the causal cone at scale $z$ is comprised of a hypercubic block of linear size $l_z$ (that is, it is made of $(l_z)^D$ sites), then the causal cone at the larger scale $(z+1)$ will involve a hypercubic block of linear size $l_{z+1}$, where $l_z$ and $l_z+1$ again follow the bounds of Eq. \ref{eq:s2e5}.

\subsection{Computation of reduced density matrices} \label{sect:density}

Let us now review how to compute the reduced density matrix
\begin{equation}
\rho \equiv \textrm{tr}_{\mathcal B-\mathcal L} \left( \ket{\psi} \bra{\psi} \right) ~~~~~~~~\label{eq:s2e6}
\end{equation}
for a hypercubic block $\mathcal{B}$ of sites\cite{Vidal08}. First, we re-iterate that $\rho$ can be expressed as a tensor network that only contains the tensors in the causal cone $\mathcal{C}(\mathcal{B})$, see Fig. \ref{fig:1DCausal}. Then, $\rho_0 \equiv \rho$ can be obtained through a sequence of reduced density matrices $\rho_z$, supported on $l_z$ effective sites inside the causal cone $\mathcal{C}(\mathcal{B})$, where as before $l_z$ is the linear size of the causal cone at scale $z$.

\begin{figure}[!tbh]
\begin{center}
\includegraphics[width=8cm]{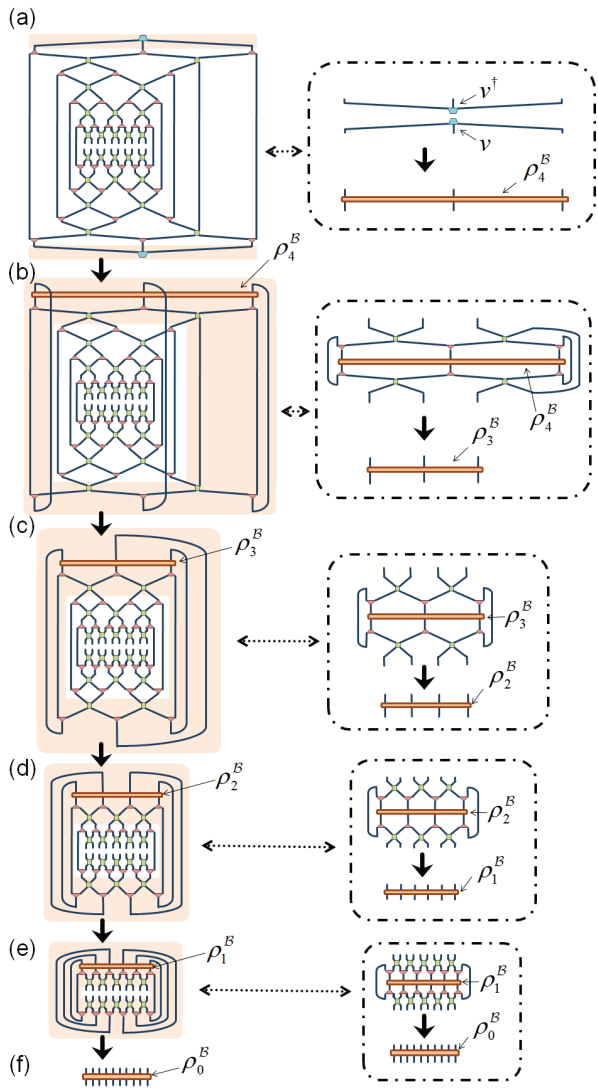}
\caption{(a) Tensor network representing the reduced density matrix $\rho$ for a block $\mathcal B$ of $10$ sites after simplification (see Fig. \ref{fig:1DCausal}). (a-f) The tensor network is contracted, resulting in $\rho_0^{\mathcal B}$, by means of a sequence of steps. Each step, depicted by a dashed box, involves applying a descending superoperator $\mathcal D$ to an intermediate density matrix $\rho_{z}^{\mathcal B}$, see Eq.\ref{eq:s2e7}.} \label{fig:DensityLower}
\end{center}
\end{figure}

Specifically, the density matrix $\rho_{z}$ at scale $z < Z$ is obtained from the density matrix $\rho_{z+1}$ at the larger scale $z+1$ through the application of a linear map called descending superoperator $\mathcal{D}$,
\begin{equation}
\rho_{z+1} \mathop \to \limits^{\mathcal{D}}  \rho_{z}, \label{eq:s2e7}
\end{equation}
where the descending superoperator $\mathcal D$ is in turn a small tensor network involving disentanglers $u$, and isometries $w$ at scale $z$, see Fig. \ref{fig:DensityLower}. The density matrix $\rho_{Z}$ at the largest length scale $z=Z$ can be obtained directly from the top tensor $v$ of the MERA, see Fig. \ref{fig:DensityLower}(a). Thus the density matrix $\rho_{0}$ is obtained at the end of a sequence of density matrices that descend through the causal cone,
\begin{equation}
    \rho_{Z} \rightarrow  \rho_{Z-1} \rightarrow \cdots \rightarrow \rho_{2} \rightarrow \rho_{1} \rightarrow \rho_{0}, \label{eq:s2e8}
\end{equation}
as depicted in Fig. \ref{fig:DensityLower}. In the next section we will use this sequence to derive an upper bound to the entanglement entropy of $\rho_0$.

\section{Scaling of entanglement entropy in the MERA} \label{sect:entscale}

In this section we reproduce the derivation of an upper bound for the scaling of entanglement entropy in the MERA presented in Ref. \onlinecite{Vidal08} (see also Ref. \onlinecite{Barthel10}). Here we provide a number of details that go beyond those provided in the original derivation.

\subsection{Entropic upper bounds}

Let us start by setting a general framework, which we will use both for the MERA in this section and for the branching MERA in Sect. \ref{sect:branchbound}.

We first recall two simple and well-known results. The von Neumann entropy of a $m \times m$ matrix representing a density matrix $\rho$ is upper bounded by
\begin{equation}
S(\rho) \leqslant \log_2\left( m \right). \label{eq:s3a}
\end{equation}
Second, if $\rho^{AB}$ denotes the density matrix of two sites with vector spaces of dimension $\chi^A$ and $\chi^B$, then the entropy of $\rho^A \equiv \tr_B [\rho^{AB}]$ is upper bounded by
\begin{equation}
S(\rho^A) \leqslant S(\rho^{AB}) + \log_2\left( \chi^B \right). \label{eq:s3b}
\end{equation}
Eq. \ref{eq:s3b} follows from the triangular inequality $S(\rho^{AB}) > |S(\rho^A) - S(\rho^{B})|$, see Ref. \onlinecite{Nielsen}.

In particular, Eq. \ref{eq:s3a} implies that any reduced density matrix $\rho_z$ in the sequence of Eq. \ref{eq:s2e8} is upper bounded by
\begin{equation}
S(\rho_z) \leq (l_z)^D \log_2\left( \chi \right), \label{eq:s3e1}
\end{equation}
since $\rho_z$ is supported on $(l_z)^D$ sites, each represented by a vector space of dimension $\chi$, and it is therefore an $m\times m$ matrix with $m = \chi^{(l_z)^{D}}$. In turn, Eq. \ref{eq:s3b} implies that
\begin{equation}
S\left( \rho_{z} \right) \leqslant S\left( \rho_{z+1} \right) + \log_2 \left( \chi \right) n^{\tra}_{z}, \label{eq:s3e2}
\end{equation}
where $n^{\tra}_{z}$ is the number of sites traced out in going from $\rho_{z+1}$ to $\rho_{z}$ by means of the linear map $\mathcal{D}$, Eq. \ref{eq:s2e7}. That is, at each step of the sequence of density matrices in Eq. \ref{eq:s2e8}, the entropy can at most increase by an amount proportional to the number $n^{\tra}_{z}$ of the sites that are being traced out.

We can now apply Eq. \ref{eq:s3e2} recursively $z'$ times starting from $z=0$ to obtain an upper bound on the entropy of $\rho_{0}$ at scale $z=0$,
\begin{equation}
S( \rho_0 ) \leqslant {\log _2}(\chi)\left( (l_{z'})^D + N^{\tra}_{z'} \right), \label{eq:s3e3}
\end{equation}
where
\begin{equation}
~~~N^{\tra}_{z'} \equiv \sum\limits_{z = 0}^{z' - 1} n^{\tra}_{z}.\label{eq:s3e3sec}
\end{equation}
This upper bound is made of two contributions. The first contribution is proportional to $(l_{z'})^D$, and corresponds to the entropy of $\rho_{z'}$ at scale $z'$. The second term is proportional to the cumulative number $N^{\tra}_{z'}$ of sites that are traced out in transforming $\rho_{z'}$ into $\rho_0$. For each choice of $z'$ in the interval $0 \geq z' \geq Z$  we obtain a different upper bound. From now on we assume that the lattice $\mathcal{L}$ is infinite, so that $Z \rightarrow \infty$ and $z'$ can be any positive integer. Below we will consider the specific upper bound obtained by setting $z'$ in Eq. \ref{eq:s3e3} to be the crossover scale $\zz$ at which the causal cone enters its stationary regime. In appendix \ref{sect:derivation} we will show that the resulting upper bound is optimal, in that it is, to leading order in $l_0$, the tightest upper bound we could obtain from Eq. \ref{eq:s3e3}.

\begin{figure}[!tbh]
\begin{center}
\includegraphics[width=8cm]{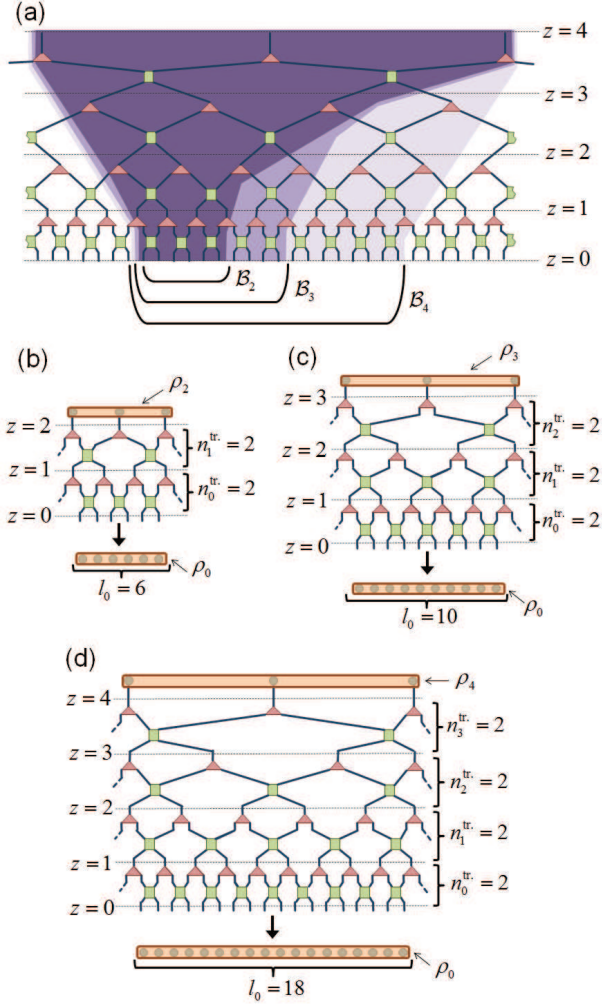}
\caption{(a) Blocks $\mathcal B_{\zz}$ of length $l_0 = 2^{\zz} + 2$ for $\zz=2,3,4$, with causal cones $\mathcal C(\mathcal B_{\zz})$ shaded, that have been chosen at special locations such that only $n^{\tra}_{z}=2$ sites are traced out at any level $z$ in obtaining density matrix $\rho_{z}$ from $\rho_{z+1}$ for all $z \leq \zz$, as illustrated (b) for block $\mathcal B_2$ (c) for block $\mathcal B_3$ and (d) for block $\mathcal B_4$.} \label{fig:1DBlock}
\end{center}
\end{figure}

\subsection{Choice of block $\mathcal{B}$} \label{sect:block}

The next step is to make a particular choice of hypercubic block $\mathcal{B}$. We restrict our attention to a block of linear size $l_0\equiv l$ given by
\begin{equation}\label{eq:s3xx}
    l_{0} = {2^{\zz}} + 2
\end{equation}
for a positive integer $\zz$, which is taken at a special location (with respect to the tensor network) that minimizes the number of sites that must be traced out in order to obtain the density matrix $\rho_{0} \equiv \rho$. This special location is such that, for length scales $z < \zz$, which correspond to the shrinking regime, no disentanglers straddle the boundary of the causal cone from the previous level, see Fig. \ref{fig:1DBlock} for examples in $D=1$ dimensions. This has two implications.

First, in the shrinking regime $z < \zz$, the width $l_z$ of the causal cone changes with the scale $z$ as
\begin{equation}\label{eq:s3xx2}
    l_{z+1}=(l_{z}+2)/2,
\end{equation}
which is a special case of Eq.\ref{eq:s2e5}. This leads to
\begin{equation}
l_{z}=\left\{\begin{array}{cc}
        \frac{l_{0}-2}{2^{z}} + 2 & \mbox{for } z < \zz \\
        3 & \mbox{for } z \geq \zz
      \end{array}
      \right.
\label{eq:s3e4}
\end{equation}
It thus follows that the scale
\begin{equation}
 \zz = \log _2(l_0 - 2) \label{eq:s3e5}
\end{equation}
is the crossover scale connecting the shrinking regime ($z < \zz$) and the stationary regime ($z\geq \zz$).

\begin{figure}[!tbh]
\begin{center}
\includegraphics[width=7cm]{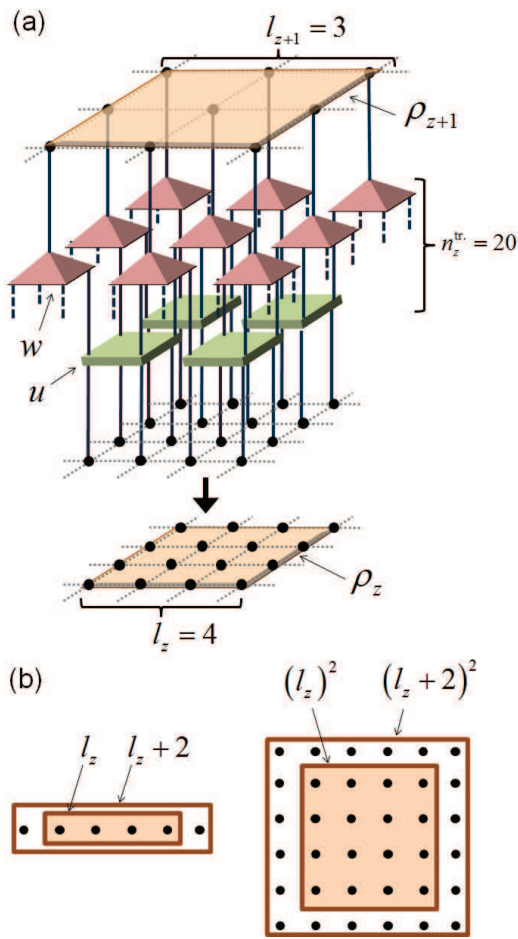}
\caption{(a) In a MERA in $D=2$ dimensions, the density matrix $\rho_{z}$, defined on a block of $( l_{z} )^2=16$ sites, is obtained by descending density matrix $\rho_{z+1}$, defined on a block of $( l_{z+1} )^2=9$ sites, tracing out $n_{z}^\textrm{tr.}=(l_{z}+2)^2-(l_{z})^2 = 6^2-4^2 = 20$ sites in the process. 
(b) Illustration of Eq. \ref{eq:nTra1} for $D=1$ and $D=2$.
} \label{fig:2DBlock}
\end{center}
\end{figure}

Second, also in the shrinking regime, the number $n^{\tra}_z$ of sites that are traced out in computing $\rho_z$ from $\rho_{z+1}$ is minimal, and given by
\begin{equation}
n^{\tra}_z = (l_z + 2)^D - (l_z)^{D}. \label{eq:nTra1}
\end{equation}
The meaning of this expression is illustrated in Fig. \ref{fig:2DBlock}(b) for $D=1,2$ dimensions, and it is very intuitive: $n^{\tra}_z$ is proportional to the size of the boundary of the causal cone $\mathcal{C}(\mathcal{B})$ at scale $z$,
\begin{equation}
n^{\tra}_{z} \approx 2D{\left( {{l_{z}}} \right)^{D - 1}} \label{eq:nTra2} 
\end{equation}
where $\approx$ indicates that we may have neglected subleading terms of order $(l_z)^{D-2}$. 

Let us then evaluate the upper bound of Eq. \ref{eq:s3e3} for the above choice of block $\mathcal{B}$, Eqs. \ref{eq:s3e4} and \ref{eq:nTra1}, and setting $z'$ as the crossover scale $\zz$ in Eq. \ref{eq:s3e5}. We obtain
\begin{equation}
S( \rho_0 ) \leqslant {\log _2}(\chi)\left( 3^D + N^{\tra}_{\zz} \right), \label{eq:s3e3b}
\end{equation}
where
\begin{equation}
~~~N^{\tra}_{\zz} = \sum\limits_{z = 0}^{\zz - 1} \left( (l_z + 2)^D - (l_z)^{D} \right).\label{eq:s3e3bsec}
\end{equation}

Next we evaluate these expressions for $D=1$ dimensions and for $D>1$ dimensions.

\subsection{Scaling in $D=1$ space dimensions} \label{sect:1DMERA}

Let us first use Eqs. \ref{eq:s3e3b}-\ref{eq:s3e3bsec} to obtain an upper bound for the entropy $S_l$ of a block of $l_{0}$ contiguous sites in $D=1$ dimensions. This result was originally derived in Ref. \onlinecite{Vidal08b} (see also Ref. \onlinecite{Barthel10}).

In this case Eq. \ref{eq:nTra1} reduces to
\begin{equation}
n^{\tra}_z = 2, \label{eq:nTra3}
\end{equation}
that is, at each length scale the same number of sites are traced out. Therefore the total number $N^{\tra}_{\zz}$ of sites that are traced out in the shrinking regime of the causal cone is proportional to the number of length scales $\zz$ present in the shrinking regime
\begin{equation}\label{eq:s3xx4}
    N^{\tra}_{\zz} = \sum_{z=0}^{\zz - 1} 2 = 2 \zz = 2 \log_2(l_0 -2),
\end{equation}
which when replaced in Eq. \ref{eq:s3e3b} leads to
\begin{equation}
S( \rho_0 ) \leqslant {\log _2}(\chi)\left( 3 + 2\log_2(l_0 -2) \right). \label{eq:s3aaa}
\end{equation}
Thus, to leading order in $l_0\equiv l$, we obtain the following upper bound for the scaling of entanglement entropy in the MERA in $D=1$ dimensions:
\begin{equation}
S_l \leqslant k_1 \log_2(l), \label{eq:s3e8}
\end{equation}
where $k_1$ is a constant that depends only on the bond dimension $\chi$. This upper bound is in agreement with a large body of numerical work\cite{Vidal03b, Pfeifer09, Evenbly10}.

\subsection{Scaling in $D>1$ space dimensions} \label{DdimMERA}

Let us now use Eqs. \ref{eq:s3e3b}-\ref{eq:s3e3bsec} to obtain an upper bound for the entropy $S_l$ of a hypercubic block of linear size $l_{0}$ in $D>1$ dimensions. This result was originally derived in Ref. \onlinecite{Vidal08b} (see also Ref. \onlinecite{Barthel10}).

In this case $n^{\tra}_z$ of Eq. \ref{eq:nTra1} reads
\begin{equation}
n^{\tra}_{z} \approx 2D \left(\frac{l_0}{2^z} \right)^{D-1} \label{eq:nTra4}
\end{equation}
where $\approx$ indicates that we have neglected subleading terms of order $(l_0)^{D-2}$ in . To leading order in $l_0$, we have
\begin{eqnarray}
N^{\tra}_{\zz} &\approx& 2D (l_0)^{D-1} \sum\limits_{z = 0}^{\zz-1} (\frac{1}{2^{D-1}})^z  \label{eq:Sum1}\\
&\leqslant& 2D (l_0)^{D-1}\sum\limits_{z = 0}^\infty (\frac{1}{2^{D-1}})^z,  \label{eq:Sum3} \\
&=& \left( \frac{2D}{1 - 2^{1-D}} \right) (l_0)^{D - 1}. \label{eq:Sum4}
\end{eqnarray}
Note that in going from Eq. \ref{eq:Sum1} to Eq. \ref{eq:Sum3} the finite geometric series was replaced by an infinite geometric series, which was then explicitly summed to give Eq. \ref{eq:Sum4}. When replaced in Eq. \ref{eq:s3e3b}, this leads to
\begin{equation}
S(\rho_0) \leqslant \log _2 (\chi) \left( 3^D + \left( \frac{2D}{1 - 2^{1-D}} \right) (l_0)^{D - 1} \right),
\end{equation}
Thus, to leading order in $l_0\equiv l$, we obtain the following upper bound for the scaling of entanglement entropy in the MERA in $D>1$ dimensions:
\begin{equation}
S_l \leqslant k_D ~ l^{D - 1}, \label{eq:s3e8b}
\end{equation}
where $k_D$ is a constant that depends only on the bond dimension $\chi$ and dimension $D$. This upper bound is in agreement with numerical work\cite{Evenbly10}.

\subsection{Comparison of results for $D=1$ and $D>1$.}

The upper bound derived for the MERA in $D=1$ dimensions, which reproduces a logarithmic correction to the boundary law, is fundamentally different to the upper bound derived for the MERA in $D>1$ dimensions, which reproduces a strict boundary law. This difference can be understood to arise from the number of indices $n^{\tra}_z$ traced out in computing the density matrix $\rho_z$ from $\rho_{z+1}$, Eq. \ref{eq:nTra1}. 

The case of $D=1$ dimensions is special in that $n^{\tra}_{z}$ is a constant as a function of the scale $z$, Eq. \ref{eq:nTra3}, meaning that all scales in the shrinking regime of the causal cone contribute \textit{equally} to the entanglement entropy of the block. Since there are $O(\log_2(l))$ such scales, we obtain the logarithmic scaling. Instead, in $D>1$ dimensions, $n^{\tra}_{z}$ decays exponentially with $z$, Eq. \ref{eq:nTra4}, and as a result the scaling of entanglement entropy is already dominated by the contribution at scale $z=0$, which is proportional to the boundary of the system, $n_0^{\tra} \approx 2D(l)^{D-1}$, thus leading to the boundary law.

The above discussion points out at how to reproduce a logarithmic correction to the boundary law in $D>1$ dimensions. Indeed, what we need is a generalization of the MERA in $D>1$ dimensions such that $n_z^{\tra}$ is constant (and proportional to $l^{D-1}$) throughout the entire shrinking regime of the causal cone, as it occurs in $D=1$ dimensions. The branching MERA, discussed next, can accomplish precisely this.

\begin{figure}[!tbh]
\begin{center}
\includegraphics[width=8cm]{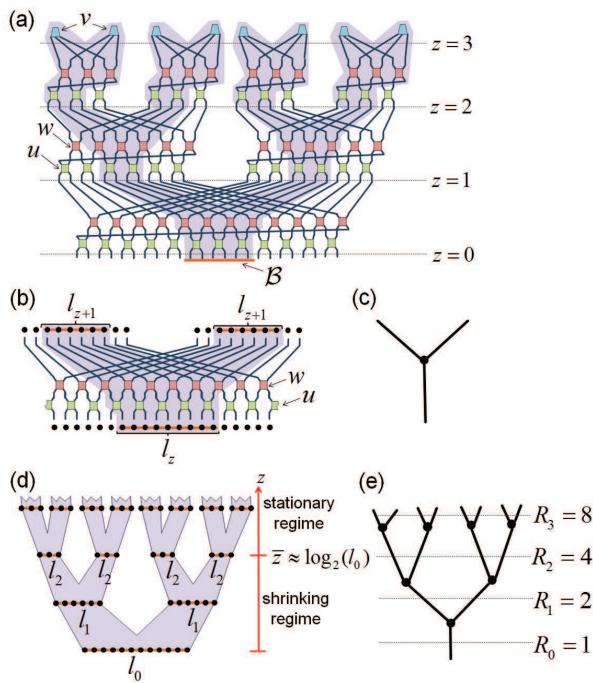}
\caption{
(a) A branching MERA, comprised of disentanglers $u$, decouplers $w$ and top tensors $v$, for a lattice of $N=24$ sites in $D=1$ dimensions. This branching MERA is based on a holographic tree with branching ratio $b=2$, such that it has $R_z = 2^z$ distinct branches at scale $z$. The causal cone $\mathcal C (\mathcal B)$ (shaded region) of block $\mathcal B$ includes tensors on all branches.
(b) An illustration of the causal cone of a block $\mathcal B$ under a single layer of the branching MERA. Disentanglers $u$ enlarge the linear size of the causal cone by at most a constant. Decouplers $w$ both enlarge the linear size of the causal cone and then split it into two branches, each containing an equal number of sites.
(c) Holographic tree representing a single instance of branching. (d) Schematic representation of the causal cone of a block of sites in a branching MERA in $D=1$ dimensions based upon a holographic tree with branching ratio $b=2$, which involves $l_z$ sites on each of the $2^z$ branches at scale $z$. (e) Holographic tree for the branching MERA in (a).} \label{fig:1DBranchCausal}
\end{center}
\end{figure}

\section{Fundamentals of the branching MERA}
\label{sect:fundbranch}

Here we review some background material on the branching MERA\cite{Evenbly12a,Evenbly12b} that is required in order to derive an upper bound on the scaling of entanglement entropy. Specifically, we briefly review the notion of holographic tree, the structure of causal cones, and the computation of reduced density matrices. These topics were discussed in depth in Ref. \onlinecite{Evenbly12b}, to which we refer the interested reader for more details.

The branching MERA is a tensor network for quantum states on a lattice in $D$ dimensions. It generalizes the MERA, in that the MERA is recovered as the particular case (namely, of trivial holographic tree, as discussed below). A main motivation for the introduction of the branching MERA is that it is capable of displaying corrections to the boundary law for entanglement entropy in $D>1$ dimensions, where both PEPS and MERA are restricted to obeying a strict boundary law.

\subsection{Holographic trees}

Both the MERA and the branching MERA are based on coarse-graining transformations that follow the same principle: the use of  local disentanglers to remove short-range entanglement. However, while in the MERA the coarse-graining transformation produces a single coarse-grained many-body systems, in the branching MERA the coarse-graining transformation produces two or more coarse-grained many-body systems. Under iteration of the coarse-graining transformation to larger scales $z$, more and more coarse-grained many-body systems are produced. The so-called holographic tree\cite{Evenbly12a,Evenbly12b} is used to specify the branching structure of the coarse-grained many-body systems produced as a function of the scale $z$.

An interesting subclass of holographic trees are those with a regular branching structure, such that each node has exactly $b$ child nodes, or uniform branching ratio $b$. A regular holographic tree with branching ratio $b=2$, or binary tree, and the corresponding branching MERA in $D=1$ dimensions are depicted in Fig. \ref{fig:1DBranchCausal}. For the purposes of this paper, a useful characterization of a holographic tree is through the total number of branches at scale $z$, which we denote $R_z$. Notice that a regular tree with branching ratio $b$ has exactly $R_z = b^z$ branches at depth $z$.

The maximal possible branching, compatible with the isometric constraints required on the tensors of the tensor network, is $b=2^D$, in which case all tensors are unitary. On the other extreme we have a tree with trivial branching, $b=1$, that is, with just one branch, which corresponds to the MERA. We will see later that an intermediate choice, $b=2^{D-1}$, leads to a branching MERA that reproduces a logarithmic correction to the boundary law in any dimension $D$.

\subsection{Causal cones} \label{sect:branchcausal}

The causal cone $\mathcal C(\mathcal B)$ for a block $\mathcal B$ of sites is defined in the branching MERA in the same way as in the MERA, namely as the set of tensors within the tensor network that are involved in the computation of the reduced density matrix $\rho$ on this block.

A causal cone in the branching MERA inherits the branching structure dictated by the holographic tree\cite{Evenbly12b}. As a result, at scale $z$ the causal cone consists of $R_z$ decoupled pieces, one for each branch, see Fig. \ref{fig:1DBranchCausal}. The linear size $l_z$ of each branch at scale $z$ depends on the scale $z$ and on the linear size $l_0\equiv l$ of the block $\mathcal{B}$ in the same way as in the MERA. In particular, one again finds the shrinking and stationary regimes discussed in Sect. \ref{sect:causal}: the causal cone of a hypercubic block of linear size $l_0 \gg 1$ shrinks to some minimum width at a crossover length scale $\zz \approx \log_2 (l_0)$, after which it remains in the stationary regime, see Fig. \ref{fig:1DBranchCausal}(d).

\subsection{Computation of reduced density matrices}

\begin{figure}[!tbh]
\begin{center}
\includegraphics[width=6cm]{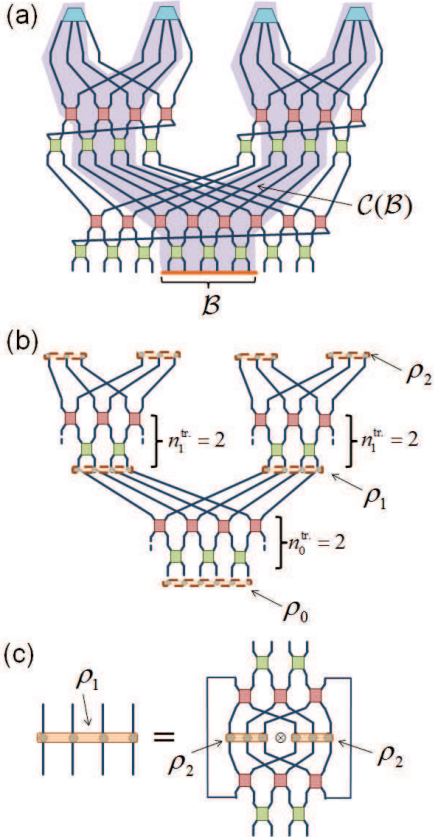}
\caption{(a) A branching MERA defined on a lattice of $N=16$ sites in $D=1$ dimensions, with the causal cone $\mathcal C (\mathcal B)$ of the block of sites $\mathcal B$ shaded. (b) The reduced density matrix $\rho_0$ supported on $\mathcal B$ is obtained by descending several copies of the reduced density matrices $\rho_z$ through the causal cone. (c) Density matrix $\rho_1$ is obtained from a tensor network involving a tensor product of two copies of $\rho_2$, as a special case of Eq.\ref{eq:s4e2}.} \label{fig:1DBranchBlock}
\end{center}
\end{figure}

At scale $z$, the reduced density matrix on the causal cone $\mathcal{C}(\mathcal{B})$ is the tensor product of $R_z$ reduced density matrices, one for each of the branches at that scale\cite{Evenbly12b}. From now on, to ease the notation we assume that these reduced density matrices are equal, so that the overall state in the causal cone at scale $z$ is simply $(\rho_z)^{\otimes R_z}$, where $\rho_z$ is the reduced density matrix on a single branch of the causal cone. [This assumption does not affect the resulting upper bound on entanglement entropy.] Then the density matrix $\rho_z$ can be computed from a tensor network involving the local reduced density matrices $\rho_{z+1}$ on each of its child branches at depth $z+1$. If there are $b$ such child branches, then
\begin{equation}
(\rho _{z + 1})^{\otimes b} \mathop\to\limits^\mathcal{D} \rho _z, \label{eq:s4e2}
\end{equation}
where $\mathcal D$ represents a descending superoperator for the branching MERA, see Fig. \ref{fig:1DBranchBlock}(c) for an example. Thus, as in the MERA, the density matrix $\rho_0$ for region $\mathcal{B}$ of the physical lattice $\mathcal{L}$ is obtained at the end of a sequence of density matrices that descend through the causal cone $\mathcal{C}(\mathcal{B})$,
\begin{equation}
    \left( \rho_Z \right)^{\otimes R_Z} \rightarrow \left( \rho_{Z-1}\right)^{\otimes R_{Z-1}} \rightarrow \cdots \rightarrow \left( \rho_1\right)^{\otimes R_1} \rightarrow \rho_0, \label{eq:s4e1}
\end{equation}
see Fig. \ref{fig:1DBranchBlock}(b). In the next section we will use this sequence to derive an upper bound to the entanglement entropy of $\rho_0$.

\section{Scaling of entanglement entropy in the branching MERA}
\label{sect:branchbound}

In this section we derive an upper bound for the scaling of entanglement entropy in the branching MERA. This derivation generalizes that of Sect. \ref{sect:entscale} in the presence of a non-trivial branching tree.

\subsection{Entropic upper bounds}

The computation of the local reduced density matrix $\rho_z$ in branching MERA, as described by Eq. \ref{eq:s4e2}, implies replacing the upper bound on entanglement entropy of Eq. \ref{eq:s3e2} with the new upper bound
\begin{equation}
S(\rho _z) \leqslant b S\left( {{\rho _{z + 1}}} \right) + {\log _2}\left( \chi  \right) n_z^{\tra}, \label{eq:s4e2b}
\end{equation}
where $n_z^{\tra}$ is the number of sites traced out in applying the descending superoperator $\mathcal D$ in Eq. \ref{eq:s3e2}. That is, the entropy of $\rho_z$ cannot be more than the sum of the entropies of the $b$ copies of $\rho_{z+1}$ it is obtained from, plus the entropy potentially gained in tracing out $n^{\tra}_{z}$ sites. 

This bound can be applied recursively $z'$ times starting at $z=0$ to obtain an upper bound for the entropy of $\rho_0$ at scale $z=0$,
\begin{equation}
S(\rho_0) \leqslant \log _2(\chi) \left( R_{z'} (l_{z'})^D + N^{\tra}_{z'} \right), \label{eq:s4e3}
\end{equation}
where
\begin{equation}\label{eq:s4e3x}
    N^{\tra}_{z'} \equiv \sum\limits_{z = 0}^{{z'}  - 1} {{R_z}n_z^{\tra}}.
\end{equation}
This upper bound is made of two contributions. The first contribution is proportional to $R_{z'}(l_{z'})^D$ and corresponds to the entropy of the $R_{z'}$ copies of $\rho_{z'}$ at scale $z'$. The second term is proportional to the cumulative number $N^{\tra}_{z'}$ of sites that are traced out in transforming $(\rho_{z'})^{\otimes N_{z'}}$ into $\rho_0$, and now includes contributions from all the branches at scale $z'$. For each choice of $z'$ in the interval $0 \geq z' \geq Z$  we again obtain a different upper bound. From now on we assume that the lattice $\mathcal{L}$ is infinite, so that $Z \rightarrow \infty$ and $z'$ can be any positive integer.

\subsection{Choice of block $\mathcal{B}$} \label{sect:blockBranch}

In order to proceed further, we evaluate Eqs. \ref{eq:s4e3} and \ref{eq:s4e3x} after making the particular choice of hypercubic block $\mathcal{B}$ described in Sect. \ref{sect:block}, see Eqs. \ref{eq:s3xx} - \ref{eq:nTra2}, and choosing the scale $z'$ to be the crossover scale $\zz$. We obtain
\begin{equation}
S(\rho_0) \leqslant \log _2(\chi) \left( R_{\zz} 3^D + N^{\tra}_{\zz} \right), \label{eq:s4aa1}
\end{equation}
where
\begin{eqnarray}\label{eq:s4aa2}
    N^{\tra}_{\zz} &\equiv& \sum\limits_{z = 0}^{\zz  - 1} R_z \left( (l_z + 2)^D -(l_z)^D\right) \label{eq:s4bb1}\\
     &\approx& 2D (l_0)^{D-1} \sum\limits_{z = 0}^{\zz-1} R_z\left(\frac{1}{2^{D-1}}\right)^z  \label{eq:s4bb2}\\
    &=& 2D(l_0)^{D-1} f(l_0) \label{eq:s4bb3}
\end{eqnarray}
where in Eq. \ref{eq:s4bb2} we have only kept leading order in $l_0$, and where
\begin{equation}
f(l_0) \equiv \sum\limits_{z = 0}^{\zz - 1} R_z \left( \frac{1}{2^{D-1}} \right)^z. \label{eq:s4e6}
\end{equation}
Thus we see that $N_{\zz}^{\tra}$ scales as the boundary law ${\left( {{l_0}} \right)^{D - 1}}$ times a multiplicative correction $f(l_0)$ that depends on the branching structure of the underlying holographic tree through $R_z$.
It follows that the entanglement entropy $S(\rho_0)$ is bounded above by
\begin{equation}
S(\rho_0) \leqslant k_D {\left( {{l_0}} \right)^{D - 1}}f\left( {{l_0}} \right), \label{eq:s4e7}
\end{equation}
where the constant $k_D$ depends on $\chi$ and $D$ (but is independent of $l_0$). Here we have used that the first term on the rhs of Eq. \ref{eq:s4aa1}, which also depends on $l_0$ through $R_{\zz}$, can be seen to be of subleading order in $l_0$, when compared to $(l_0)^{D-1}f(l_0)$, for any relevant choice of $R_{z}$.

Next we evaluate function $f(l_0)$ for two classes of holographic trees.

\begin{figure}[!tbh]
\begin{center}
\includegraphics[width=8cm]{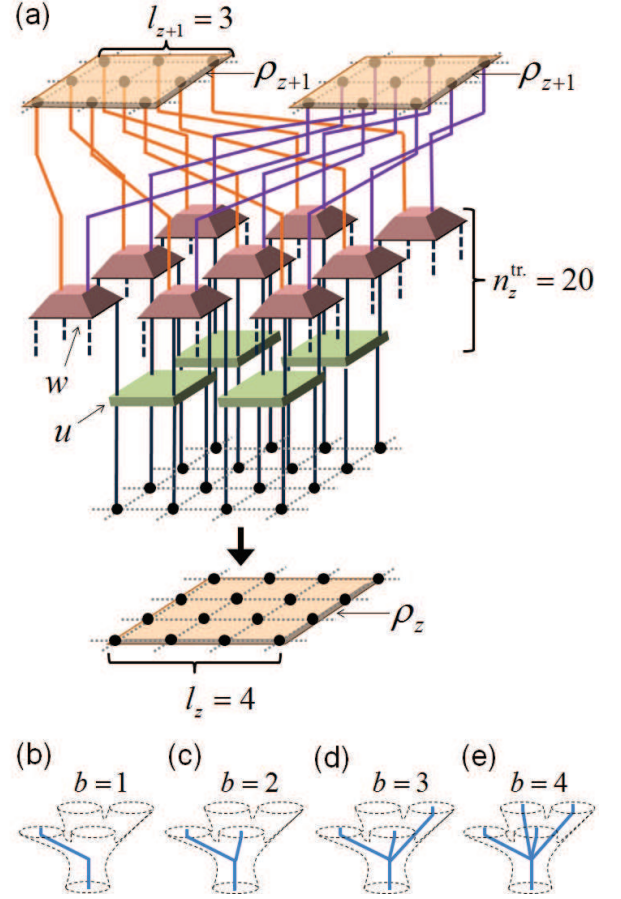}
\caption{(a) A depiction of part of a branching MERA in $D=2$ dimensions. The density matrix $\rho_{z}$ is obtained by combining two copies of $\rho_{z+1}$ with isometries/decouplers $w$ and disentanglers $u$, and then tracing out $n_{z}^{\tra}=20$ indices. (b-e) A branch of the branching MERA in $D=2$ dimensions can split into $b=1,2,3,4$ sub-branches at each level. Diagram (a) corresponds to the case of $b=2$.} \label{fig:2DBranchBlock}
\end{center}
\end{figure}

\subsection{Regular holographic trees}

Let us evaluate the above upper bound on entanglement entropy for branching MERA with a regular holographic tree with branching ratio $b$, where each node of the tree has exactly $b$ child nodes. Notice that for this family of trees the number of branches at depth $z$ scales as $R_z=b^z$. Then the function $f(l_0)$ of Eq.\ref{eq:s4e6}, which describes the multiplicative correction to the boundary law, becomes
\begin{equation} \label{eq:s4ddd}
f\left( {{l_0}} \right) = \sum\limits_{z = 0}^{\zz  - 1} \left( \frac{b}{2^{D-1}}\right)^z.
\end{equation}
Notice that this is a geometric series with common ratio $r=b{2^{1 - D}}$ and, as such, can be summed explicitly. This sum takes has a different functional dependance on $l_0$ contingent on whether the branching $b$ is such that the common ratio is greater than, equal to or less-than unity. In these three cases, to leading order in $l_0$ the function $f(l_0)$ reads
\begin{equation}
f\left( {{l_0}} \right) \approx \left\{ {\begin{array}{*{20}{l}}
  {{c_1}}&{b < {2^{D - 1}}} \\
  {{c_2}{{\log }_2}({l_0})}&{b = {2^{D - 1}}} \\
  {{c_3}{{\left( {{l_0}} \right)}^{\left( {1 - D + {{\log }_2}(b)} \right)}}}&{b > {2^{D - 1}}}
\end{array}} \right.
\end{equation}
for some constants $c_1$, $c_2$, and $c_3$ that depend on $D$ and $b$ (but are independent of $l_0$). These, together with Eq. \ref{eq:s4e7}, lead to the following upper bounds for the scaling of entanglement in the branching MERA with a regular holographic tree
\begin{equation}
S_l \leq \left\{ {\begin{array}{*{20}{l}}
  \tilde{c}_1 ~l^{D-1}&{b < {2^{D - 1}}} \\
  \tilde{c}_2 ~l^{D-1} \log_2(l) &{b = {2^{D - 1}}} \\
  \tilde{c}_3 ~l^{\log_2(b)} &{b > {2^{D - 1}}}
\end{array}} \right.
\end{equation}
for some constants $\tilde{c}_{\alpha} = c_{\alpha} k_D$ that depend on $D$, $b$, and $\chi$. A subset of these results can be found on table \ref{table:BranchMERA}. Notice in particular that for $b=2^{D-1}$ we obtain a logarithmic correction to the boundary law for all dimensions $D$, whereas $b=2^{D}$ produces a bulk law.

\subsection{Beyond regular holographic trees} \label{sect:Beyond}

\begin{figure}[!tbh]
\begin{center}
\includegraphics[width=8cm]{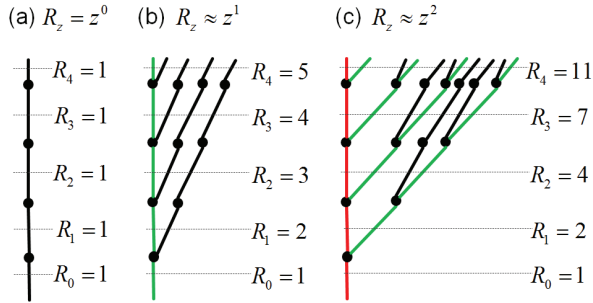}
\caption{(a) Un-branched holographic tree corresponding to the MERA. (b) Holographic tree whose number of branches scales linearly with depth, $R_z = z$. (c) Holographic tree whose number of branches scales quadratically with depth, $R_z = z^2/2 +z/2 +1$.} \label{fig:IrregTree}
\end{center}
\end{figure}

Branching MERA can be of course also based upon holographic trees other than regular trees with a homogeneous branching ratio $b$, which implies $R_z=b^z$. As an example, here we consider a second family of branching MERA where the number $R_z$ of branches scales with the scale $z$ as $(2^{D-1})^z$ (as is the case of a regular tree with $b=2^{D-1}$) but corrected by a power of $z$,
\begin{equation}
{R_z} = (2^{D-1})^z{z^\kappa },
\end{equation}
for a positive integer $\kappa$. The function $f$ of Eq.\ref{eq:s4e6} that describe multiplicative corrections to the boundary law evaluates as,
\begin{align}
  f\left( {{l_0}} \right) &= \sum\limits_{z = 0}^{\xi  - 1} {{z^\kappa }}  \hfill \nonumber \\
   &\leqslant \int\limits_{z = 0}^{\xi  - 1} {{z^\kappa }dz}  \hfill \nonumber \\
   &\leqslant {c_4}{\left( {{{\log }_2}({l_0})} \right)^{\kappa  + 1}} \hfill.
\end{align}
Thus the entanglement entropy in this branching MERA is upper bounded by polylogarithmic multiplicative corrections to the boundary law, 
\begin{equation}\label{eq:s4ccc}
    S_l \lesssim \tilde{c}_4 l^{D-1} \left( \log_2 (l) \right)^{\kappa+1},
\end{equation}
for some constant $\tilde{c}_4 = c_4k_D$. Fig. \ref{fig:IrregTree} shows specific holographic trees that lead to such upper bounds for leading power $\kappa=0,1,2$ in $D=1$ dimensions. More generally, branching MERA with other forms of non-regular holographic trees are expected to lead to other exotic scaling of entanglement entropy.

\section{Saturation of the upper bounds: a numerical study} \label{sect:numresults}

In Sects. \ref{sect:entscale} and \ref{sect:branchbound} we derived upper bounds for the scaling of entanglement entropy both in the MERA and in several instances of branching MERA, respectively. These upper bounds suggested various forms of scaling of the entanglement entropy $S_l$ as a function of the linear size $l$ of a hypercubic block of sites. However, these derivations do not imply that such forms of scaling are actually realized in practice. To show that they are, we turn now to numerics.

Specifically, the upper bounds that we have found can be generically written as
\begin{equation}\label{eq:generic}
    S_l \leq k_D~ l^{D-1} ~f(l), 
\end{equation}
where $f(l)$ is some function [e.g. a constant, a (poly-)logarithm, or a power of $l$] that measures departure from the boundary law and where only the leading order in $l$ has been considered. The constant $k_D$ depends on the number $D$ of spatial dimensions and is proportional to $\log_2(\chi)$.

Notice that we do not expect that the actual scaling of entanglement entropy obtained numerically will match $k_D$, because this constant resulted from assuming that every time that an index was traced out (see Eqs. \ref{eq:s3e2} and \ref{eq:s4e2b}) this added the maximal possible amount of entropy $\log_2(\chi)$ to the density matrix, whereas in practice one expects a smaller amount only bounded by $\log_2(\chi)$. Here, what we would like to confirm numerically is that our upper bounds for $S_l$ capture the proper (leading order) functional dependence in $l$ [that is, confirm that $f(l)$ indeed scales e.g. as a constant, a (poly-)logarithm, or a power of $l$].

Two more remarks are in order. First, it is always possible to come up with specific choices of the variational parameters that e.g. make any form of (branching) MERA obey a strict boundary law and thus have trivial $f(l)$, by using the tensor network to exactly represent a shortly correlated state. However, here we are interested in the scaling $S_l$ of a given tensor network for a generic choice of variational parameters. In practice we will consider a homogeneous tensor network, in which a unit cell of two tensors (e.g. one disentangler $u$ and one isometry $w$ in the MERA) loaded with random coefficients is repeated throughout the network.

\begin{figure}[!tbph]
\begin{center}
\includegraphics[width=8cm]{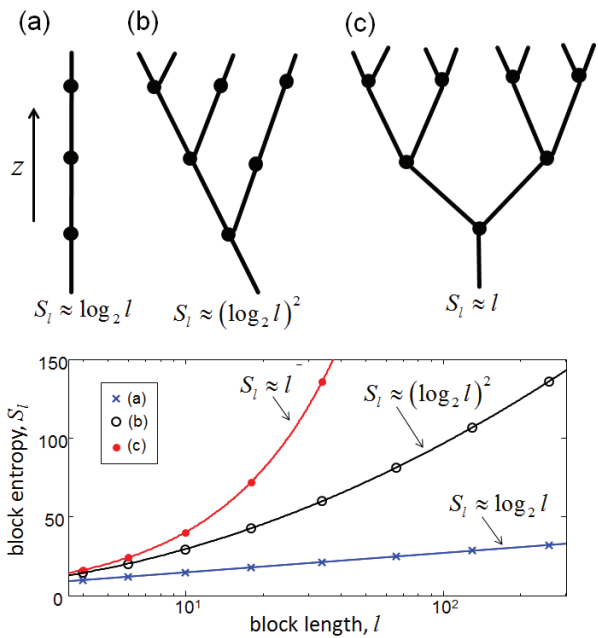}
\caption{(top) Holographic trees for (a) a regular MERA, (b) a branching MERA where the number of branches increases linearly with depth and (c) a branching MERA where the number of branches increases exponentially with depth, together with the predicted scalings of entanglement entropy on $1D$ lattices. (bottom) Entanglement entropy $S_l$ for blocks of length $l$ computed numerically from randomly initialized (branching) MERA in $D=1$ dimensions with the holographic trees shown in (a), (b), and (c), together with fits over the indicated function form.} \label{fig:1DNumeric}
\end{center}
\end{figure}

Second, a density matrix $\rho$ supported on $l^D$ sites is a matrix whose dimensions grow exponentially in $l$, and the computation of its entropy $S_l$ requires an effort which is also exponential in $l$ (with some tricks, this effort can be made instead roughly exponential in $S_l$). As a result, only small values of $l$ (respectively $S_l$) can be considered, which makes it hard to numerically confirm the various forms of scaling suggested by the analytical upper bounds. To overcome this difficulty, here we use the free fermion formalism, which lowers the computational cost to being just the third power of the number $l^{D}$ of sites. The price we pay is that we will choose the variational parameters randomly within a restricted subset of parameters, such that the tensors can be obtained by exponentiating a quadratic form of the fermionic creation and annihilation operators\cite{Evenbly10}. While this clearly constraints the variational parameters we consider, there is no obvious reason why such restriction should affect the scaling of entanglement entropy of the resulting network and, as a matter of fact, there is plenty of numerical work in $D=1$ dimensions that confirms that the scaling $S_l$ observed in the MERA for free fermions is the same as for interacting systems.

\begin{figure}[!tbph]
\begin{center}
\includegraphics[width=8cm]{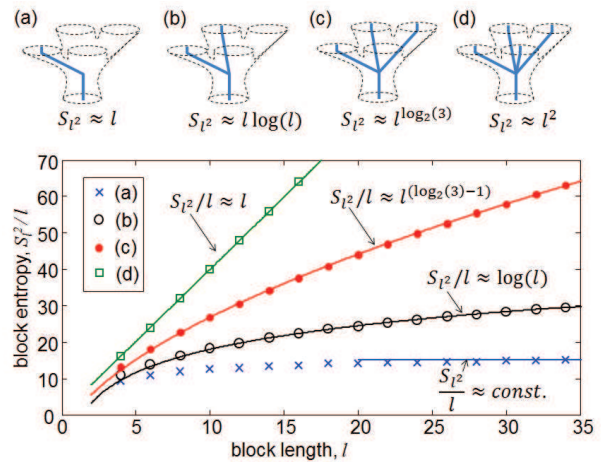}
\caption{(top) (a)-(d) Predicted scalings of entanglement entropy for the branching MERA in $D=2$ dimensions based upon regular holographic trees with branching ratio $b=1,2,3,4$. (bottom) Entanglement entropy $S_{l^2} / l$ for a square block of linear length $l$ computed numerically from randomly initialized tensors, for a branching MERA with the holographic trees shown in (a)-(d), together with fits over the indicated function form.} \label{fig:2DNumeric}
\end{center}
\end{figure}

Figs. \ref{fig:1DNumeric} and \ref{fig:2DNumeric} show the scaling of entanglement entropy for several instances of (branching) MERA in $D=1$ dimensions and $D=2$ dimensions, respectively. As mentioned above, in each case a pair of tensors with random coefficients (compatible with the isometric constraints and within the free fermion formalism) were used throughout the entire tensor network. The scaling displayed in these two figures was typical over many choices of random coefficients, and shows agreement with the upper bounds we have derived in Sects. \ref{sect:entscale} and \ref{sect:branchbound}. In particular, it confirms the ability to obtain (poly-)logarithmic and power-law corrections to the boundary law by adjusting the holographic tree of the branching MERA.

\section{Conclusions} \label{sect:conclusion}

In this paper we have reviewed the scaling of entanglement entropy in the MERA (derived in Ref. \onlinecite{Vidal08}) and established a number of forms of scaling of entanglement in the branching MERA. We have both provided upper bounds for this scaling and, in several cases, numerical confirmation that the upper bounds are saturated to leading order by using free fermion systems.

The upper bounds are based on the examination of how one can actually compute reduced density matrices in the (branching) MERA [see also appendix \ref{sect:geoscale}], and suggest a scaling of the form
\begin{equation}\label{eq:concl1}
    S_l \approx l^{D-1} f(l),
\end{equation}
for various choices of the correction $f(l)$ to the boundary law $l^{D-1}$, including poly-logarithmic corrections
\begin{equation}\label{eq:concl2}
    f(l) = (\log(l))^{\kappa}
\end{equation}
for positive integers $\kappa$; and polynomial corrections
\begin{equation}\label{eq:concl2b}
    f(l) = l^{\alpha}
\end{equation}
for $0\leq \alpha \leq 1$. The particular choice of $f(l)$ depends on details of the holographic tree that characterizes the pattern of branching in the branching MERA.

Perhaps the most relevant construction corresponds to a regular tree with branching ratio $b=2^{D-1}$, which reproduces the logarithmic correction to the boundary law, Eq. \ref{eq:s1scale2}, observed in the ground state of certain highly entangled phases of matter, such as Fermi liquids and spin-Bose metals\cite{Motrunich07}. It should be clearly noted, however, that reproducing the proper scaling of entanglement entropy is not in itself a sufficient condition for the branching MERA to be a good variational ansatz for the ground state of such systems. This is indeed a subject that requires further investigation. Nevertheless, we report that preliminary studies based on free fermions seem to indicate that indeed certain ground states in $D=2$ dimensions with a $D_{\Gamma}=1$ Fermi surface are well represented with a branching MERA.

The present investigation of entanglement entropy in the branching MERA has revealed forms of scaling, such as $f(l)\approx (\log_2(l))^2$, that have not yet been found in known phases of matter. An intriguing question is then whether it might be possible to engineer local Hamiltonians such that their ground states obey such exotic scalings. The structure of the branching MERA, and the interpretation of the holographic tree as describing decoupling into several subsystems at large length scale/low energies, actually give important clues about how one could go about engineering such Hamiltonians. As a matter of fact, through the study of the branching MERA with the holographic tree depicted in Fig. \ref{fig:IrregTree}(b), it is possible to design a free fermion Hamiltonian with algebraic decay of interactions such that its ground state displays a log-squared violation of the boundary law, $S_{l}\sim \left(\log_2(l)\right)^2$, in $D=1$ dimensions\cite{Evenbly13}. More generally, we envisage that the branching MERA will provide a useful formalism to investigate and design other exotic states of quantum matter.


\appendix

\newpage
\section{Optimal upper bounds from Eqs. \ref{eq:s4e3}-\ref{eq:s4e3x} } \label{sect:derivation}

In this appendix we investigate the optimality of the upper bounds for entanglement entropy derived in Sects. \ref{sect:entscale} and \ref{sect:branchbound}. In particular, whether they offer the tightest upper bound that could possibly be derived from Eqs. \ref{eq:s4e3}-\ref{eq:s4e3x}, which we rewrite here:
\begin{equation}
S(\rho_0) \leqslant \log _2(\chi) \left( R_{z'} (l_{z'})^D + N^{\tra}_{z'} \right), \label{eq:As4e3}
\end{equation}
and
\begin{equation}\label{eq:As4e3x}
    N^{\tra}_{z'} \equiv \sum\limits_{z = 0}^{{z'}  - 1} {{R_z}n_z^{\tra}}.
\end{equation} 
By tightest possible upper bound we mean an upper bound that already captures the most restrictive function $f(l)$ in Eq. \ref{eq:s4e7} (to leading order in $l$ and up to a multiplicative constant). For instance, in Sects. \ref{sect:entscale} and \ref{sect:branchbound} we have obtained that, for some (branching) MERA, $f(l)$ scales as $f(l)\approx \log_2(l)$. Here we investigate if it could have been possible to use Eqs. \ref{eq:As4e3}-\ref{eq:As4e3x} to instead obtain a more restrictive functionality for $f(l)$, such as a constant $f(l)$ in this example.

Recall that, for any value $z'\geq 0$, Eqs. \ref{eq:As4e3}-\ref{eq:As4e3x} provide an upper bound for the entanglement entropy of a particular choice of hypercubic block $\mathcal{B}$ of sites, introduced in Sect. \ref{sect:block} and characterized by Eqs. \ref{eq:s3xx}-\ref{eq:nTra2}. Let us reproduce here some of these equations. The linear size $l_0$ of the block is chosen to be
\begin{equation}\label{eq:s8xx}
    l_{0} = {2^{\zz}} + 2
\end{equation}
for some positive integer $\zz$. As a function of scale $z$, the linear size of the causal cone reads
\begin{equation}
l_{z}=\left\{\begin{array}{cc}
        \frac{l_{0}-2}{2^{z}} + 2 & \mbox{for } z < \zz \\
        3 & \mbox{for } z \geq \zz
      \end{array}
      \right.
\label{eq:s8e4}
\end{equation}
so that $\zz$ denotes the scale at which the shrinking regime ($z < \zz$) and stationary regime ($z\geq \zz$) of the causal cone meet. In the shrinking regime, the number $n^{\tra}_z$ of sites that are traced out in going from scale $z+1$ to scale $z$ is given by
\begin{equation}
n^{\tra}_z = (l_z + 2)^D - (l_z)^{D}. \label{eq:s8nTra1}
\end{equation}

In Sects. \ref{sect:entscale} and \ref{sect:branchbound} we have made a particular choice of $z'$ in Eqs. \ref{eq:As4e3}-\ref{eq:As4e3x}, namely $z'=\zz$, with
\begin{equation}\label{eq:Across}
    \zz \equiv \log_2(l_0-2),
\end{equation}
that is, at the scale where the causal cone achieves its smallest linear size, $l_{\zz} = 3$. We expected this upper bound to be optimal, in the sense specified above. In Sect. \ref{sect:numresults}, this expectation has been confirmed numerically for several forms of (branching) MERA.

Let us then justify analytically that another choice of $z'$ in Eqs. \ref{eq:As4e3}-\ref{eq:As4e3x}, $z' \neq \zz$, could not have produce a sharper upper bound to leading order. For simplicity, below we restrict our attention to a regular holographic tree. However, similar arguments can be applied for a branching MERA with a more complicated branching structure.

Let us first rewrite Eqs. \ref{eq:As4e3}-\ref{eq:As4e3x} as
\begin{equation}\label{eq:apA1}
 S(l_0) \leq \log_2(\chi) F(z'),
\end{equation}
where
\begin{equation} \label{eq:apA2}
    F(z') \equiv R_{z'}(l_{z'})^D + \sum_{z=0}^{z'-1} R_{z}[(l_{z}+2)^D - (l_{z})^D],
\end{equation}
and where we have used Eq. \ref{eq:s8nTra1}. Let us also define
\begin{eqnarray}
    \Delta(z') &\equiv& F(z'+1) - F(z')  \nonumber \\
    &=& R_{z'+1}(l_{z'})^D - R_{z'}(l_{z'})^D + R_{z'}[(l_{z'}+2)^D - (l_z')^D]  \nonumber\\
    &=& R_{z'} \left[ \left(1+\frac{R_{z'+1}}{2^D R_{z'}} \right)(l_{z'}+2)^D - 2(l_{z'})^D \right], \label{eq:apA3}
\end{eqnarray}
which measures how $F(z')$ changes in increasing the scale from $z'$ to $z'+1$. If $\Delta(z')$ is positive (negative), then scale $z'$ provides a tighter (respectively, looser) upper bound than scale $z'+1$. For a regular holographic tree with branching ratio $b$ $(1\leq b \leq 2^D)$, such that at scale $z$ there are $R_z = b^z$ branches, we have
\begin{equation}\label{eq:apA4}
\Delta(z') =b^{z'} \left[ \left(1+\frac{b}{2^D} \right)(l_{z'}+2)^D - 2(l_{z'})^D \right].
\end{equation}
Below we argue that, with one exception (also discussed), $\Delta(z')$ is always negative for a sufficiently large linear size $l_{z^*}$ independent of $l_0$. This will be seen to imply that for any fixed $b$, the upper bound that we have obtained in Sect. \ref{sect:branchbound} by setting $z'= \zz$ in Eqs. \ref{eq:As4e3}-\ref{eq:As4e3x} and the upper bounds that we would have obtained by setting instead the optimal scale $z' = z^{*}$ are essentially equivalent.

Let us first consider the exception, which occurs when we have the maximal allowed branching ratio $b=2^D$ (unitary limit of the tensor network). Then we have
\begin{equation}\label{eq:Aexept}
\Delta(z') = 2^{Dz'}\left[ 2(l_{z'}+2)^D - 2(l_{z'})^D \right] > 0,
\end{equation}
which implies that the tightest upper bound is already obtained for the choice $z'=0$, which leads to the bulk law $S_l \approx l^D$. 

Let us now consider any other allowed branching ratio $b < 2^D$. In this case, $\Delta(z')$ in Eq. \ref{eq:apA4} is negative for sufficiently large values of $l_{z'}$, because
\begin{equation}\label{eq:A5}
    1+\frac{b}{2D} < 2
\end{equation}
and therefore for sufficiently large $l_{z'}$ the term proportional to $(l_{z'})^D$ always beats the term proportional to $(l_{z'}+2)^D$. On the other hand, for $z'=\zz$, which corresponds to $l_{z'}=3$, we have
\begin{equation}\label{eq:apA4b}
\Delta(\zz) =b^{\zz} \left[ \left(1+\frac{b}{2^D} \right)(5)^D - 2(3)^D \right],
\end{equation}
which can be seen to be positive for any $D$ and valid $b$.
In particular, it can be seen that $\Delta(z')$ only changes sign at some finite $l_{z^*}$ that does not depend on $l_0$. That is, $l_{z^*}$ occurs at a scale $z^*$ such that $\zz-z^*$ is independent of $l_0$. Then, setting $z^*$ instead of $\zz$ in the geometric series of Eq. \ref{eq:s4ddd} only results in eliminating a constant number of terms, (corresponding to large length scales $z$ from $z^*$ to $\zz$; or, equivalently, small linear sizes $l_z$ from $l_{z^*}$ to $l_{\zz}=3$) from that sum. For sufficiently large $l_0$, these terms add to a contribution that is essentially independent of $l_{0}$, and therefore have no consequences for the leading order scaling of $f(l_0)$.

\begin{figure}[!tbh]
\begin{center}
\includegraphics[width=8cm]{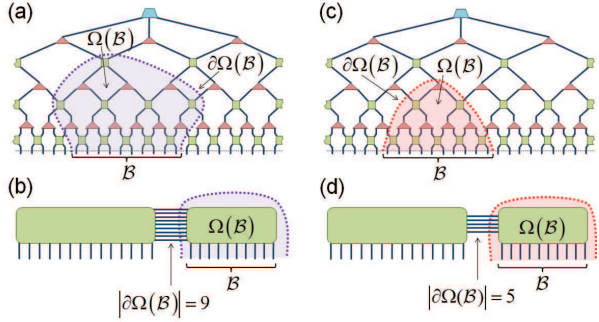}
\caption{(a) For a block of sites $\mathcal B$, the MERA is divided into two parts by identifying a region $\Omega(\mathcal B)$ containing the open indices corresponding to $\mathcal B$. The network in (a) is simplified by contracting all tensors in $\Omega(\mathcal B)$ down to a single tensor (and likewise for its complement). The size of the boundary, $\left| {\partial \Omega (\mathcal{B})} \right|=9$, can be used to bound the entanglement entropy of region $\mathcal B$. (c) An alternative choice of region $\Omega(\mathcal B)$ containing the open indices corresponding to $\mathcal B$. (d) This alternative choice of region has a boundary $\left| {\partial \Omega (\mathcal{B})} \right|=5$, and thus gives a tighter entropic bound on $\mathcal B$ than the choice of region from (a).} \label{fig:HoloDef}
\end{center}
\end{figure}

\section{Geometric upper bound for the scaling of entanglement entropy} \label{sect:geoscale}

In this paper we have discussed upper bounds for the entanglement entropy in the (branching) MERA that result from considering the number of indices that are traced out in computing a density matrix, Eq.  \ref{eq:s4e3}-\ref{eq:s4e3x}. On the other hand, another way of obtaining an upper bound for the entanglement entropy of a region $\mathcal{B}$ from a tensor network representing the state $\ket{\Psi}$ of the whole system is by counting the number of bond indices that need to be cut in order to split the tensor network into two pieces, one corresponding to region $\mathcal{B}$ and another corresponding to the rest of the system. This has been extensively discussed e.g. in Refs. \onlinecite{Swingle12,Evenbly11a}. This second approach, purely geometrical, does not require that the tensors in the network fulfill isometric constraints (nor, by extension, the presence of well-defined causal cones), and it is hence applicable to any class of tensor network state. In addition, it connects naturally with holographic calculations of entanglement entropy\cite{Nishioka09}.

Specifically, let $\Omega(\mathcal{B})$ be a region of the tensor network that contains the physical indices corresponding to the sites in $\mathcal{B}$ and no other physical index, see Fig. \ref{fig:HoloDef}. Then the number of bond indices crossing the boundary $\partial \Omega(\mathcal{B})$ of $\Omega(\mathcal{B})$, denoted $|\partial \Omega (\mathcal{B})|$ and referred to as the \textit{size of the boundary of} $\Omega(\mathcal{B})$, provides an upper bound for the entropy of the reduced density matrix $\rho$ in region $\mathcal{B}$,
\begin{equation}
S(\rho) \le \log_2 (\chi) \left| {\partial \Omega (\mathcal{B})} \right|. \label{eq:A1}
\end{equation}
Notice that, for a given block $\mathcal B$, there will be many different ways to choose the region $\Omega (\mathcal{B})$. Different choices of $\Omega (\mathcal{B})$ may have different boundary sizes $\left| \partial \Omega (\mathcal{B}) \right|$, thus giving different upper bounds to the entropy. The tightest upper bound comes from the minimally connected region $\Omega (\mathcal{B})$, i.e. that with the smallest size $\left| \partial \Omega (\mathcal{B}) \right|$ of the boundary $\partial \Omega (\mathcal{B})$.

\begin{figure}[!tbh]
\begin{center}
\includegraphics[width=8cm]{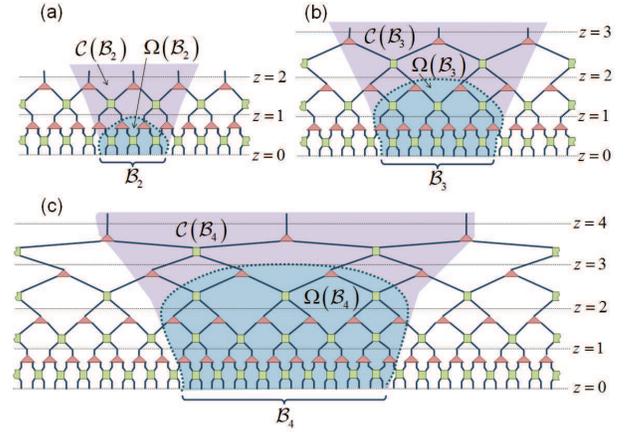}
\caption{Blocks of length $l=2^{\zz} +2$ for (a) $\zz=2$, (b) $\zz=3$, and (c) $\zz=4$, showing causal cones $\mathcal{C}(\mathcal{B}_{\zz})$ of these blocks together with the minimally connected holographic regions $\Omega(\mathcal{B}_{\zz})$. Notice $\mathcal C(\mathcal B_{\zz})$ and $\Omega(\mathcal{B}_{\zz})$ are exactly coincident for depths $z\le \zz-2$.} \label{fig:HoloEquivLarge}
\end{center}
\end{figure}

Is there a relationship between the upper bounds obtained in this paper and the geometric upper bounds that one can obtain in the branching MERA? Notice that the strategy followed in this paper can be recast in geometric terms. Indeed, the causal cone $\mathcal{C}(\mathcal{B})$ is an example of region $\Omega(\mathcal{B})$, and the indices that are traced out in computing the density matrix $\rho$ for region $\mathcal{B}$ correspond to its boundary. More specifically, in this paper we considered the sites that were traced out in the shrinking regime of the causal cone. Denoting by $\mathcal{C}^{\shr}(\mathcal{B})$ this part of the causal cone, our upper bound for the entanglement entropy can be re-expressed as
\begin{equation}\label{eq:A1b}
    S_l \leq \log_2 (\chi) |\partial \mathcal{C}^{\shr}(\mathcal{B})|.
\end{equation}

It is then natural to ask whether the geometric upper bound obtained by following the causal cone in its shrinking regime is already optimal, or one could find a tighter geometric upper bound. A difficulty in answering this question is that we do not know how to systematically find minimally connected surfaces in the MERA. However, it is plausible that in the shrinking regime a causal cone indeed defines a minimally connected surface\cite{Swingle12}, except for small changes at the top of $\partial \mathcal{C}^{\shr}(\mathcal{B})$ which only introduce sub-leading corrections to the scaling. For instance, Fig. \ref{fig:HoloEquivLarge} depicts what we believe to be minimally connected surfaces in the $D=1$ MERA. These indeed correspond to the boundary of the causal cone in the shrinking regime, and only depart from it when we are approaching the stationary regime at scales near $\zz = \log_2(l-2)$, where $l$ is the linear size of the region $\mathcal{B}$. As a result, the optimal geometric upper bound and the upper bounds that we have obtained in this paper indeed only seem to differ in some minor details and scale, to leading order, in the same way with $l$. We believe that this picture holds for the MERA in any dimension $D$, and that it may also hold in the branching MERA.


\begin{thebibliography}{99}
\bibitem{Other}
For each choice of lattice $\mathcal{L}$, one can consider several different implementations of the MERA, in correspondence with the different ways of coarse-graining the lattice compatible with entanglement renormalization (removal of short-range entanglement while preserving the structure of locality. These implementations may differ in a number of aspects, including how many sites are coarse-grained into a single site, or the detailed computational efficiency, but are essentially equivalent. In particular, the resulting MERAs obey the same scaling of entanglement entropy.
\end{thebibliography}
\end{document}